\documentclass[prl,10pt,twocolumn,showpacs,preprintnumbers,amsmath,amssymb,aps]{revtex4}

\usepackage{graphicx}
\usepackage{dcolumn}
\usepackage{bm}
\usepackage{amssymb}
\usepackage{amsmath}
\usepackage{amsthm}
\usepackage{yhmath}

\font\SYM=msbm10
\newcommand{\Real}{\mbox{\SYM R}}

\begin{document}

\preprint{}

\title{Geometric invariant measuring the deviation from Kerr data}

\author{Thomas B\"ackdahl}
 \email{t.backdahl@qmul.ac.uk}
\author{Juan A. Valiente Kroon}%
 \email{j.a.valiente-kroon@qmul.ac.uk}
\affiliation{%
School of Mathematical Sciences, Queen Mary University of London,\\
 Mile End Road, London E1 4NS, United Kingdom
}%


\date{\today}

\begin{abstract}
A geometrical invariant for regular asymptotically Euclidean data for
the vacuum Einstein field equations is constructed. This invariant
vanishes if and only if the data correspond to a slice of the Kerr
black hole spacetime ---thus, it provides a measure of the
non-Kerr-like behavior of generic data. In order to proceed with the
construction of the geometric invariant, we introduce the notion of
approximate Killing spinors.
\end{abstract}

\maketitle

\emph{Introduction.---} It is widely expected that the late time
behavior of a dynamical black hole spacetime will approach, in some
suitable sense, the Kerr spacetime. Making sense of this expectation
is one of the outstanding challenges of modern general relativity. In
particular, clarifying what it means that a spacetime is close to the
Kerr spacetime is of great relevance for the problem of the non-linear
stability of the Kerr spacetime and for the numerical evolution of
black holes. Because of the coordinate freedom in general relativity it
is, in general, difficult to measure how much two spacetimes differ from
each other. Nevertheless, invariant characterizations of spacetimes
provide a way of bridging this difficulty.

Most analytical and numerical studies of the Einstein field equations
make use of a 3+1 decomposition of the equations and the
unknowns. Thus, it is important to have a characterization of the Kerr
solution which is amenable to this type of splitting. Most known
invariant characterizations of the Kerr spacetime have problems in
this or other respects. For example, the characterization of the Kerr
spacetime in terms of the so-called \emph{Mars-Simon tensor} requires
the \emph{a priori} existence of a Killing vector in the spacetime
\cite{Mar99,Mar00}. An invariant characterization in terms of
concomitants of the Weyl tensor produces very involved expressions
when performing a 3+1 split \cite{FerSae09,GarVal10}. Furthermore, the
above characterizations are local by construction, and it is not clear
how they could be used to produce a global characterization of initial
data sets. In this letter we discuss an alternative characterization
of the Kerr spacetime and show how it can be used to obtain a global
geometrical invariant of asymptotically Euclidean slices of a
spacetime. This geometric invariant has the key property of vanishing
if and only if the hypersurface is a slice of the Kerr spacetime. In
this sense, our invariant is analogous to the invariant characterising
time symmetric slices of static spacetimes discussed in \cite{Dai04c}.


\smallskip
\emph{Killing spinors and Petrov type D spacetimes.---} Let
$(\mathcal{M},g_{\mu\nu})$ be an orientable and time orientable globally
hyperbolic vacuum spacetime. A Killing spinor is a symmetric spinor
$\kappa_{AB}=\kappa_{(AB)}$ satisfying
\begin{equation}
\label{Killing1} \nabla_{A'(A} \kappa_{BC)}=0,
\end{equation} where $\nabla_{AA'}$ denotes the spinorial counterpart
of the Levi-Civita connection of the metric $g_{\mu\nu}$. Here
$A,\,B,\cdots$ denote abstract spinorial indices, while 
$\mathbf{A}, \,\mathbf{B},\cdots$ will denote indices with respect to a
specific frame. The spinorial conventions of \cite{PenRin84} are
used. Killing spinors offer a way of relating properties of the
curvature with properties of the symmetries of the spacetime. Given a
Killing spinor $\kappa_{AB}$, one has that
$\xi_{AA'}=\nabla^B{}_{A'} \kappa{}_{AB}$ is a complex Killing vector
of the spacetime.

We note a local characterization of the Kerr spacetime in terms of
Killing spinors based on the following results: (i) a vacuum
spacetime admits a Killing spinor, $\kappa_{AB}$, if and only if it is of
Petrov type D, N or O \cite{Jef84,GarVal08a} (a Petrov type D
spacetime for which $\xi_{AA'}$ is real will be called a \emph{generalized Kerr-NUT
spacetime} \cite{FerSae07,DebKamMcL84});  (ii) Kerr is always of type D (there are no points where it degenerates to N
or O)  and is the only asymptotically flat generalized Kerr-NUT
spacetime \cite{Mar99,Mar00}. Let $\Psi_{ABCD}$ denote the Weyl spinor.  One has the following:

\smallskip
Theorem 1.---\emph{Let $(\mathcal{M},g_{\mu\nu})$ be an asymptotically
flat spacetime for which $\Psi_{ABCD}\neq 0$ and
$\Psi_{ABCD}\Psi^{ABCD}\neq 0$. Then $(\mathcal{M},g_{\mu\nu})$ is
isometric to the Kerr spacetime if and only if there exists a
Killing spinor such that the associated Killing vector is real.}




\medskip
\emph{Asymptotically Euclidean slices.---} Let
$(\mathcal{S},h_{ab},K_{ab})$ denote a smooth initial data set for the
vacuum Einstein field equations ---that is, $(h_{ab},K_{ab})$ satisfy
the vacuum constraint equations on $\mathcal{S}$. In what follows, the
3-manifold $\mathcal{S}$ will be assumed to be asymptotically
Euclidean with two asymptotic ends, $i_1, \;i_2$. An
asymptotic end is an open set diffeomorphic to the complement of an
open ball in $\Real^3$. The fall off conditions of the various fields
will be expressed in terms of weighted Sobolev spaces $H^s_\beta$,
where $s$ is a non-negative integer and $\beta$ is a real number. We
say that $\eta\in H^\infty_\beta$ if $\eta\in H^s_\beta$ for all
$s$. In what follows we use the theory for these spaces developed in
\cite{ChoChr81} written in the conventions of \cite{Bar86}. Thus, the
functions in $H^\infty_\beta$ are smooth over $\mathcal{S}$ and have a
fall off at infinity such that $\partial^l \eta =
o(r^{\beta-|l|})$. We will often write $\eta=o_\infty(r^\beta)$ for
$\eta\in H^\infty_\beta$ at an asymptotic end.

We assume that on each end it is possible to introduce asymptotically
Cartesian coordinates $x^i_{(k)}$, $k=1,2$, with $r=[(x^1_{(k)})^2 +
(x^2_{(k)})^2 + (x^3_{(k)})^2]^{1/2}$, such that the intrinsic metric
and extrinsic curvature of $\mathcal{S}$ satisfy
\begin{eqnarray}
&& h_{ij} = -\left(1+ 2m_{(k)}r^{-1}\right)\delta_{ij} + o_\infty(r^{-3/2}), \label{decay1} \\
&& K_{ij} = o_\infty(r^{-5/2}), \label{decay2}
\end{eqnarray}
where $i,\,j$ are coordinate indices ---in contrast to $a,\,b$ which
are taken to be abstract ones.  We assume that $m_{(k)}\geq 0$. For
simplicity we have excluded from our analysis boosted slices ---this
will be discussed elsewhere. Note, however, that the slices considered
allow a non-vanishing ADM angular momentum.

\smallskip
\emph{Killing spinor initial data.---} A set of necessary and
sufficient conditions for the development $(\mathcal{M},g_{\mu\nu})$
of the data $(\mathcal{S},h_{ab},K_{ab})$ to be endowed with a Killing
spinor was obtained in \cite{GarVal08a}. Let $\tau_{AA'}$ be the
spinor counterpart of the normal to $\mathcal{S}$, with normalization
given by $\tau_{AA'} \tau^{AA'}=2$. The spinor $\tau_{AA'}$ allows to
introduce a space spinor formalism ---see e.g. \cite{Som80,GarVal08a}
for details.  In particular, the covariant derivative $\nabla_{AA'}$
can be split according to $\nabla_{AA'} = \tfrac{1}{2} \tau_{AA'}
\nabla - \tau^Q{}_{A'}\nabla_{AQ}$, where $\nabla \equiv \tau^{AA'}
\nabla_{AA'}$ and $\nabla_{AB} \equiv \tau_{(A}{}^{A'}\nabla_{B)A'}$
is the Sen connection. The Sen connection is not intrinsic to the
hypersurface $\mathcal{S}$, however, it can be expressed in terms of
the spinorial Levi-Civita connection of $h_{ab}$, $D_{AB}$, and of the
spinorial counterpart of $K_{ab}$,
$K_{ABCD}=K_{(AB)(CD)}=K_{CDAB}$. One has, for example, that
$\nabla_{AB} \pi_C = D_{AB}\pi_C + \tfrac{1}{2} K_{ABC}{}^{D} \pi_D$.
Given a spinor $\pi_{A}$, we define its Hermitian conjugate via
$\hat{\pi}_{A} \equiv \tau_{A}{}^{E'}\bar{\pi}_{E'}$. The Hermitian
conjugate can be extended to higher valence symmetric spinors in the
obvious way.  The spinors $\nu_{AB}$ and $\xi_{ABCD}$ are said to be
real if $\hat{\nu}_{AB}=-\nu_{AB}$ and
$\hat{\xi}_{ABCD}=\xi_{ABCD}$. It can be verified that $\nu_{AB}
\hat{\nu}^{AB}, \; \xi_{ABCD} \hat{\xi}^{ABCD}\geq 0$. If the spinors
are real, then there exist real tensors $\nu_a$, $\xi_{ab}$ such that
$\nu_{AB}$ and $\xi_{ABCD}$ are their spinorial counterparts. Notice
that $\hat{D}_{AB}=-D_{AB}$. The Killing vector
$\xi_{AA'}=\nabla^B{}_{A'} \kappa_{AB}$ can be decomposed in terms of
its lapse, $\xi$, and shift, $\xi_{AB}$, according to $\xi_{AA'} =
\tfrac{1}{2} \tau_{AA'} \xi - \tau^{Q}{}_{A'}\xi_{AQ}$, where
\begin{eqnarray}
 && \xi \equiv \tau^{AA'}\xi_{AA'}=\nabla^{AB}\kappa_{AB}, \\
&& \xi_{AB} \equiv \tau_{(A}{}^{A'}
\xi_{B)A'}=\tfrac{3}{2}\nabla^{P}{}_{(A}\kappa_{B)P}.
\end{eqnarray}

Some extensive computer algebra calculations carried out in the suite
xAct \cite{xAct} show that the conditions
found in \cite{GarVal08a} for the existence of a Killing spinor in the
development of $(\mathcal{S},h_{ab},K_{ab})$ are equivalent to:
\begin{eqnarray}
&& \nabla_{(AB} \kappa_{CD)}=0,\label{kspd1}\\
&& \Psi_{(ABC}{}^F\kappa_{D)F}=0, \label{kspd2}\\
&& 3\kappa_{(A}{}^E\nabla_B{}^F\Psi_{CD)EF}+\Psi_{(ABC}{}^F\xi_{D)F}=0, \label{kspd3}
\end{eqnarray}
where $\xi_{AB}$ is used as a shorthand for
$\tfrac{3}{2}\nabla^{P}{}_{(A}\kappa_{B)P}$. The restriction of
$\Psi_{ABCD}$ to the initial hypersurface $\mathcal{S}$ can be
expressed in terms of its electric and magnetic parts as
$\Psi_{ABCD}=E_{ABCD} + \mbox{i}B_{ABCD}$, where
\begin{eqnarray}
&& \hspace{-8mm} E_{ABCD}\!=
\!\tfrac{1}{6} \Omega_{ABCD} K\!-\!\tfrac{1}{2} \Omega_{(AB}{}^{PQ} \Omega_{CD)PQ}\!-\!r_{(ABCD)},\negthinspace\negthinspace \\
&& \hspace{-8mm} B_{ABCD}\!=\! \mbox{i} D^Q{}_{(A} K_{BCD)Q},
\end{eqnarray} 
where $\Omega_{ABCD}\equiv K_{(ABCD)}$ and $K\equiv K^{AB}{}_{AB}$.  The spinor
$r_{ABCD}$ is the spinorial representation of the Ricci tensor of
$h_{ab}$. All these quantities can be computed from the initial data.
From the analysis in \cite{GarVal08a} one has the following result:

\smallskip
Theorem 2.--- \emph{The development $(\mathcal{M},g_{\mu\nu})$ of an
initial data set for the vacuum Einstein field equations,
$(\mathcal{S},h_{ab},K_{ab})$, has a Killing spinor if and only if
there exists a symmetric spinor $\kappa_{AB}$ on $\mathcal{S}$
satisfying equations \eqref{kspd1}-\eqref{kspd3}.}

\smallskip
Equations (\ref{kspd1})-(\ref{kspd3}) will be collectively referred to
as the \emph{Killing spinor initial data equations}. Equation
(\ref{kspd1}) will be called the spatial Killing spinor equation
whereas (\ref{kspd2}) and (\ref{kspd3}) will be known as the
\emph{algebraic conditions}. A solution to equations
(\ref{kspd1})-(\ref{kspd3}) will be called a \emph{Killing spinor
data}, while a solution to only equation (\ref{kspd1}) will be known
as a \emph{Killing spinor candidate}.

As a consequence of Theorem 1, equations (\ref{kspd1})-(\ref{kspd3}) are
known to have a non-trivial solution if and only if the initial data
set $(\mathcal{S},h_{ab},K_{ab})$ is data for the Kerr/Schwarzschild
spacetime. For Kerr initial data satisfying the asymptotic conditions
(\ref{decay1})-(\ref{decay2}), one can always choose asymptotically
Cartesian coordinates $(x^1,x^2,x^3)$ and  orthonormal frames on the 
asymptotic ends such that
\begin{equation}
\kappa_{\mathbf{AB}} = \mp\frac{\sqrt{2}}{3}x_{\mathbf{AB}}\mp\frac{2\sqrt{2}m}{3r}x_{\mathbf{AB}}+ o_\infty(r^{-1/2}), \label{asymptotic_behavior}
\end{equation}
with
\begin{equation}
 \quad  x_{\mathbf{AB}} = \frac{1}{\sqrt{2}}
\left(
\begin{array}{cc}
-x^1 +\mbox{i}x^2 & x^3 \\
x^3 & x^1 +\mbox{i}x^2
\end{array}
\right).
\end{equation}
Using (\ref{asymptotic_behavior}) one finds
that $\xi = \pm\sqrt{2} + o_\infty(r^{-1/2})$,
$\xi_{\mathbf{AB}}=o_\infty(r^{-1/2})$. In other words, the Killing spinor of
the Kerr spacetime gives rise to its stationary Killing vector.

Crucially, a direct computation shows that for any initial data set satisfying
\eqref{decay1}-\eqref{decay2}, a spinor of the form
\eqref{asymptotic_behavior} satisfies
$\nabla_{(\mathbf{AB}}\kappa_{\mathbf{CD})}=o_\infty(r^{-3/2})$.  
 

\smallskip
\emph{Approximate Killing spinors.---} Equation (\ref{kspd1})
constitutes an overdetermined condition for the 3 complex components
of the spinor $\kappa_{AB}$. One would like to replace it by an equation which
always has a solution. For this, one notes that the operator defined
by the left hand side of equation (\ref{kspd1}) sends valence-2
symmetric spinors to valence-4 totally symmetric spinors. We note the
identity
\begin{align}
&\int_{\mathcal{U}} \nabla^{AB} \kappa^{CD} \hat{\xi}_{ABCD} \mbox{d}\mu - \int_{\mathcal{U}} \kappa^{AB} \widehat{\nabla^{CD} \xi_{ABCD}}\mbox{d}\mu  \label{integration:by:parts} \\ 
&+ \int_{\mathcal{U}} 2\kappa^{AB} \Omega^{CDF}{}_A\hat\xi_{BCDF}\mbox{d}\mu 
= \int_{\partial \mathcal{U}} n^{AB} \kappa^{CD} \hat{\xi}_{ABCD} \mbox{d}S, \nonumber
\end{align}
with $\mathcal{U}\subset \mathcal{S}$, and where $\mbox{d}S$ denotes
the area element of $\partial \mathcal{U}$, $n_{AB}$ its outward
pointing normal, and $\xi_{ABCD}$ is a symmetric spinor. Using
\eqref{integration:by:parts} one finds
that the formal adjoint of the spatial Killing spinor operator is
given by 
$\nabla^{AB}\xi_{ABCD}-2\Omega^{ABF}{}_{(C}\xi_{D)ABF}$. The
composition of the two operators is formally self-adjoint by
construction and renders the equation
\begin{align}
L(\kappa_{CD})\equiv{}&\nabla^{AB} \nabla_{(AB} \kappa_{CD)}-\Omega^{ABF}{}_{(C}\nabla_{|AB|}\kappa_{D)F}\nonumber
\\
&-\Omega^{ABF}{}_{(C}\nabla_{D)F}\kappa_{AB}=0.\label{elliptic} 
\end{align}
We shall call a solution, $\kappa_{AB}$, to equation (\ref{elliptic})
an \emph{approximate Killing spinor}. Clearly, any solution to the
spatial Killing equation (\ref{kspd1}) is also a solution to equation
(\ref{elliptic}). Equation (\ref{elliptic}) arises as the
Euler-Lagrange equation of the functional
\begin{equation}
J = \int_{\mathcal{S}} \nabla_{(AB} \kappa_{CD)} \widehat{\nabla^{AB} \kappa^{CD}} \mbox{d}\mu, \label{functional}
\end{equation}
where $\mbox{d}\mu$ denotes the volume element of the metric
$h_{ab}$. 

A calculation reveals that the operator defined by the left hand side
of this last equation is elliptic. Moreover, it can be verified that
under the asymptotic conditions (\ref{decay1})-(\ref{decay2}) the
operator is asymptotically homogeneous \cite{Can81,ChoChr81}. It
follows that the operator is a linear bounded operator with finite
dimensional Kernel and closed range \cite{ChoChr81,Loc81}.


We want to consider solutions to equation \eqref{elliptic} that behave
asymptotically like \eqref{asymptotic_behavior}. A lengthy
calculation which will be presented elsewhere renders the following:

\smallskip
Lemma 3. \emph{At any asymptotic end of an initial data set
  satisfying \eqref{decay1}-\eqref{decay2} there exists a
  $\kappa_{AB}$ such that  $\xi=\pm\sqrt{2} + o_\infty(r^{-1/2})$,
  $\xi_{AB}=o_\infty(r^{-1/2})$, $\kappa_{AB}=o_\infty(r^{3/2})$, and
  $\nabla_{(AB}\kappa_{CD)}=o_\infty(r^{-3/2})$.  In a specific
  asymptotic Cartesian frame and coordinates $\kappa_{\mathbf{AB}}$
  takes the form \eqref{asymptotic_behavior}. }

\smallskip
The solutions constructed in the previous lemma can be smoothly
cut off so they are zero outside the asymptotic end, and then added to
yield a real spinor $\mathring{\kappa}_{AB}$ on the entire slice such
that $\nabla_{(AB}\mathring{\kappa}_{CD)}\in H^\infty_{-3/2}$ with
asymptotic behavior (\ref{asymptotic_behavior}) at both ends. We
write the following ansatz for the solution to equation
(\ref{elliptic}):
\begin{equation}
\kappa_{AB} = \mathring{\kappa}_{AB} + \theta_{AB}, \quad \theta_{AB}\in H^\infty_{-1/2}. \label{asymptotic_elliptic}
\end{equation}

One has the following result:

\smallskip
Theorem 4. \emph{Given an asymptotically Euclidean initial data set
$(\mathcal{S},h_{ab},K_{ab})$ satisfying the asymptotic conditions
\eqref{decay1} and \eqref{decay2}, there exists a smooth unique
solution to equation \eqref{elliptic} with asymptotic behavior given
by \eqref{asymptotic_elliptic}.}

\smallskip 
\emph{Remark.---} Given the spinor $\kappa_{AB}$ obtained from Theorem
4, one has that by construction $\nabla_{(AB}\kappa_{CD)}\in
H^\infty_{-3/2}$, which because of Bartnik's conventions means that
$\nabla_{(AB}\kappa_{CD)}\in L^2$. Consequently, the functional $J$
given by \eqref{functional} evaluated at the solution $\kappa_{AB}$
given by Theorem 4 is finite.

\medskip
\emph{Proof of Theorem 4.---} Substitution of ansatz (\ref{asymptotic_elliptic}) into equation (\ref{elliptic}) renders the following equation for the spinor $\theta_{AB}$:
\begin{equation}
\label{elliptic:general}
L(\theta_{CD}) = -L(\mathring{\kappa}_{CD}).
\end{equation}
First, it is noticed that due to elliptic regularity, any $H^2_{-1/2}$
solution to the previous equation is in fact a $H^\infty_{-1/2}$
solution, so that if $\theta_{AB}$ exists, then it must be smooth
---see e.g. \cite{ChoChr81}. By construction it
follows that $ \nabla_{(AB} \mathring{\kappa}_{CD)}\in
H^\infty_{-3/2}$, so that $F_{CD}\equiv-L(
\mathring{\kappa}_{CD})\in H^\infty_{-5/2}$.

We make use of the Fredholm alternative for weighted
Sobolev spaces to discuss the existence of solutions to equation
(\ref{elliptic:general}) ---see e.g. \cite{Can81,Loc81}. In the particular case of equation
(\ref{elliptic:general}) there exists a unique $H^2_{-1/2}$ solution
if
\begin{equation}
\int_{\mathcal{S}} F_{AB} \hat{\nu}^{AB} \mbox{d}\mu=0
\end{equation}
for all $\nu_{AB}\in H^2_{-1/2}$ satisfying 
$L^*(\nu_{CD})=L(\nu_{CD})=0$. 
It will be shown in the sequel that such $\nu_{AB}$ must be
trivial. Using the identity (\ref{integration:by:parts}) with
$\xi_{ABCD}= \nabla_{(AB} \nu_{CD)}$ and assuming that
$L(\nu_{CD})=0$, one obtains
\begin{eqnarray}
&&\int_{\mathcal{S}} \nabla^{AB}\nu^{CD} \widehat{\nabla_{(AB} \nu_{CD)}} \mbox{d}\mu \nonumber \\
&& \hspace{1cm} = \int_{\partial\mathcal{S}_\infty} n^{AB}\nu^{CD}  \widehat{\nabla_{(AB}\nu_{CD)}} \mbox{d}S,
\end{eqnarray}
where $\partial S_\infty$ denotes the sphere at infinity. As
$\nu_{AB}\in H^2_{-1/2}$ by assumption, it follows that $\nabla_{(AB}
\nu_{CD)} \in H^\infty_{-3/2}$ and furthermore that $n^{AB} \nu^{CD}
\widehat{\nabla_{(AB}\nu_{CD)}} = o(r^{-2})$.  An integral over a finite
sphere will then be of type $o(1)$. Thus, the integral over $\partial
S_\infty$ vanishes.  Consequently,
\begin{equation}
\int_{\mathcal{S}} \nabla^{AB}\nu^{CD} \widehat{\nabla_{(AB} \nu_{CD)}} \mbox{d}\mu =0.
\end{equation}
Therefore one concludes that $\nabla_{(AB} \nu_{CD)}=0$. That is,
$\nu_{AB}$ has to be a Killing spinor candidate. Using the methods
devised in \cite{ChrOMu81} to prove that there are no non-trivial
Killing vectors of a 3-dimensional manifold that go to zero at
infinity, one can prove that if $\nu_{AB}\in H^\infty_{-1/2}$ is a
solution to the spatial Killing spinor equation (\ref{kspd1}) then
$\nu_{AB}\equiv 0$ on $\mathcal{S}$. The proof of this last result
relies on the fact that
\begin{equation}
\nabla_{AB}\nabla_{CD}\nabla_{EF} \nu_{GH}= H_{ABCDEFGH},
\end{equation} 
where $H_{ABCDEFGH}$ is a homogeneous expression of $\nu_{AB}$,
$\nabla_{AB}\nu_{CD}$ and $\nabla_{AB}\nabla_{CD}\nu_{EF}$ ---this
expression is obtained out of a lengthy computer algebra
calculation. Consequently, the Kernel of equation (\ref{elliptic})
with decay in $H^2_{-1/2}$ is trivial. Accordingly, the Fredholm
alternative imposes no restriction. Thus, there exists a unique
solution to equation (\ref{elliptic}) with asymptotic decay given by
(\ref{asymptotic_elliptic}). This completes the proof of Theorem 4.


\smallskip
\emph{The geometric invariant.---} We use the functional
(\ref{functional}) and the algebraic conditions (\ref{kspd2}) and
(\ref{kspd3}) to construct the geometric invariant measuring the
deviation of $(\mathcal{S},h_{ab},K_{ab})$ from Kerr
initial data. To this end, let $\kappa_{AB}$ be a solution to equation
(\ref{elliptic}) as given by Theorem 4, and furthermore, let $\xi_{AB}
\equiv \tfrac{3}{2}\nabla^{P}{}_{(A}\kappa_{B)P}$. Define
\begin{align}
I_1 \equiv{}& \int_{\mathcal{S}} \Psi_{(ABC}{}^{F}\kappa_{D)F}
\hat{\Psi}^{ABCG}\hat{\kappa}^{D}{}_G \mbox{d}\mu, \label{I1} \\
I_2 \equiv{}& \int_{\mathcal{S}} \left(3\kappa_{(A}{}^{E}\nabla_{B}{}^{F}\Psi_{CD)EF}+\Psi_{(ABC}{}^{F}\xi_{D)F}\right) \nonumber \\
&\times \left(3\hat\kappa^{AP}\widehat{\nabla^{BQ}\Psi^{CD}{}_{PQ}}+\hat\Psi^{ABCP}\hat\xi^{D}{}_P\right){} \mbox{d}\mu. \label{I2}
\end{align}
The geometric invariant is then defined by
\begin{eqnarray}
I \equiv J + I_1 + I_2. \label{geometric:invariant}
\end{eqnarray}
By construction $I$ is coordinate independent. From the form of the metric
(\ref{decay1}) we have $\Psi_{ABCD}\in H^\infty_{-3+\varepsilon}$,
$\varepsilon>0$.  By the multiplication lemma in \cite{ChoChr81} and
$\kappa_{AB}\in H^\infty_{1+\varepsilon}$ we have
$\Psi_{(ABC}{}^{F}\kappa_{D)F} \in H^\infty_{-3/2}$.  Thus, again one
finds that $I_1<\infty$.  A similar argument shows $I_2 <
\infty$. Hence, the invariant (\ref{geometric:invariant}) is finite
and well defined. Clearly $I\geq 0$. Note that the invariants $I_1$ and $I_2$
are not connected to a variational principle as in the case of $J$. This
is an important difference with the construction of \cite{Dai04c}.

Because of our smoothness assumptions, if $I=0$ it follows that equations
\eqref{kspd1}-\eqref{kspd3} are satisfied on the whole of
$\mathcal{S}$. Thus, the development of $(\mathcal{S},h_{ab},K_{ab})$
is, at least in a slab, of Petrov type D, N or O. The types N and O
can be excluded by requiring $\Psi_{ABCD}\neq 0$,
$\Psi_{ABCD}\Psi^{ABCD}\neq 0$ everywhere on $\mathcal{S}$. Finally,
if $I=0$ one has that the pair $(\xi,\xi_{AB})$ gives rise to a
(possibly complex) spacetime Killing vector $\xi_{AA'}$. As a
consequence of our decay assumptions,
$\xi-\hat{\xi}=o_\infty(r^{-1/2})$ and
$\xi_{AB}+\hat{\xi}_{AB}=o_\infty(r^{-1/2})$, corresponding to the
imaginary part of the Killing data $(\xi,\xi_{AB})$, give rise to a
Killing vector that goes to zero at infinity. However, there are no
non-trivial Killing vectors of this type
\cite{BeiChr96,ChrOMu81}. Thus, $\xi_{AA'}$, is a real Killing
vector. Theorems 1 and 2 render our main result:

\smallskip
Theorem 5.--- \emph{Let $(\mathcal{S},h_{ab},K_{ab})$ be an
asymptotically Euclidean initial data set for the Einstein vacuum
field equations satisfying \eqref{decay1}-\eqref{decay2} such that
$\Psi_{ABCD}\neq 0$ and $\Psi_{ABCD}\Psi^{ABCD}\neq 0$ everywhere on $\mathcal{S}$. Let $I$ be the invariant defined by equations
\eqref{functional}, \eqref{I1}, \eqref{I2} and
\eqref{geometric:invariant}, where $\kappa_{AB}$ is given as the only
solution to equation \eqref{elliptic} with asymptotic behavior given
by \eqref{asymptotic_elliptic}. The invariant $I$ vanishes if
and only if $(\mathcal{S},h_{ab},K_{ab})$ is an initial data set for
the Kerr spacetime.}




\smallskip
\emph{Applications and generalizations.---} Given the invariant of
theorem 5, a natural question to be asked is how it behaves under time
evolution. Addressing this question requires an analysis of the spinor
$\nabla \kappa_{AB}$, which can be seen to satisfy an elliptic
equation similar to (\ref{elliptic}). In
this letter we have restricted our attention to asymptotically
Euclidean slices, however, a similar analysis can be carried out on
hyperboloidal and asymptotically cylindrical slices. If some type of
constancy or monotonicity property could be established, this would be
a useful tool for studying non-linear stability of the Kerr spacetime
and also in the numerical evolutions of black hole spacetimes. For
example, it could be the case that the invariant $I$ remains
constant along the leaves of a foliation of asymptotically Euclidean
slices, while monotonicity holds only if one considers a foliation
intersecting null infinity ---like in the case of the ADM and Bondi
masses.


The decay and regularity assumptions used are certainly not
optimal. Full arguments and generalizations, will be discussed elsewhere.

\begin{acknowledgments}
We thank A Garc\'{\i}a-Parrado for his help with computer
algebra calculations in the suite xAct, and M Mars and N Kamran for valuable
comments. TB is funded by the Wenner-Gren
foundations. JAVK is funded by the EPSRC.
\end{acknowledgments}




\end{document}